\documentclass[aps,prl,groupedaddress,superscriptaddress,twocolumn,notitlepage,bibnotes]{revtex4-2}

\usepackage{newpxtext,newpxmath}

\usepackage[latin1]{inputenc}
\usepackage{amsthm}
\usepackage{amssymb}
\usepackage{amsmath}
\usepackage{bbold}
\usepackage{bbm}
\usepackage[pdftex, backref=page]{hyperref}
\hypersetup{
    colorlinks=true,
    linkcolor=blue,
    citecolor=magenta,
    urlcolor=magenta
}

\usepackage{braket}
\usepackage{dsfont}
\usepackage{mathdots}
\usepackage{mathtools}
\usepackage{enumerate}
\usepackage[shortlabels]{enumitem}
\usepackage{csquotes}
\usepackage{stmaryrd}
\usepackage[cal=boondox]{mathalfa}
\usepackage{graphicx}
\usepackage{stackengine}
\usepackage{scalerel}
\usepackage{tensor}       
\usepackage{array}
\usepackage{makecell}
\newcolumntype{x}[1]{>{\centering\arraybackslash}p{#1}}
\usepackage{tikz}
\usepackage{pgfplots}
\usetikzlibrary{shapes.geometric, shapes.misc, positioning, arrows, arrows.meta, decorations.pathreplacing, decorations.pathmorphing, patterns, angles, quotes, calc}
\usepackage{booktabs}
\usepackage{xfrac}
\usepackage{siunitx}
\usepackage{centernot}
\usepackage{comment}
\usepackage{chngcntr}
\usepackage[justification=justified,format=plain]{caption}
\usepackage{subcaption}
\usepackage{ragged2e}

\newtheorem{thm}{Theorem}
\newtheorem*{thm*}{Theorem}

\newtheorem*{prop*}{Proposition}

\newtheorem*{lemma*}{Lemma}
\newtheorem{cor}[thm]{Corollary}
\newtheorem*{cor*}{Corollary}

\newtheorem*{cj*}{Conjecture}

\newtheorem*{Def*}{Definition}

\newtheorem*{question*}{Question}

\newtheorem*{problem*}{Problem}

\makeatletter
\def\thmhead@plain#1#2#3{%
  \thmname{#1}\thmnumber{\@ifnotempty{#1}{ }\@upn{#2}}%
  \thmnote{ {\the\thm@notefont#3}}}
\let\thmhead\thmhead@plain
\makeatother

\theoremstyle{definition}

\makeatletter
\newcommand{\manualifempty}[3]{%
  \edef\@tempa{#1}%
  \ifx\@tempa\@empty
    #2
  \else
    #3
  \fi
}
\makeatother

\makeatletter
\newtheoremstyle{manualstyle}
  {3pt}{3pt}{\itshape}{}{\bfseries}{.}{ }{}

\theoremstyle{manualstyle}

\newtheorem{manualthminner}{Theorem}

\newtheorem{manualpropinner}{Proposition}

\newtheorem{manuallemmainner}{Lemma}

\newtheorem{manualcorinner}{Corollary}

\makeatother

\newcommand{\bb}{\begin{equation}\begin{aligned}\hspace{0pt}}
\newcommand{\bbb}{\begin{equation*}\begin{aligned}}
\newcommand{\ee}{\end{aligned}\end{equation}}
\newcommand{\eee}{\end{aligned}\end{equation*}}

\newcommand{\ketbra}[1]{\ket{#1}\!\bra{#1}}
\newcommand{\Ketbra}[1]{\Ket{#1}\!\Bra{#1}}
\newcommand{\ketbraa}[2]{\ket{#1}\!\bra{#2}}

\renewcommand{\epsilon}{\varepsilon}

\newcommand{\dd}{\mathrm{d}}

\newcommand{\R}{\mathds{R}}

\newcommand{\Z}{\mathds{Z}}
\newcommand{\C}{\mathds{C}}

\DeclareMathOperator{\Tr}{Tr}

\DeclareMathAlphabet{\pazocal}{OMS}{zplm}{m}{n}

\DeclareMathOperator{\Id}{id}

\newcommand{\HH}{\pazocal{H}}
\newcommand{\T}{\pazocal{T}}

\newcommand{\NN}{\pazocal{N}}

\newcommand{\BC}{\pazocal{B}}

\newcommand{\ac}{\mathbf{a}}
\newcommand{\x}{\mathbf{x}}
\newcommand{\p}{\mathbf{p}}

\newcommand{\lsmatrix}{\left(\begin{smallmatrix}}
\newcommand{\rsmatrix}{\end{smallmatrix}\right)}

\stackMath

\stackMath

\makeatletter
\newcommand*\rel@kern[1]{\kern#1\dimexpr\macc@kerna}
\newcommand*\widebar[1]{%
  \begingroup
  \def\mathaccent##1##2{%
    \rel@kern{0.8}%
    \overline{\rel@kern{-0.8}\macc@nucleus\rel@kern{0.2}}%
    \rel@kern{-0.2}%
  }%
  \macc@depth\@ne
  \let\math@bgroup\@empty \let\math@egroup\macc@set@skewchar
  \mathsurround\z@ \frozen@everymath{\mathgroup\macc@group\relax}%
  \macc@set@skewchar\relax
  \let\mathaccentV\macc@nested@a
  \macc@nested@a\relax111{#1}%
  \endgroup
}

\counterwithin*{equation}{part}
\counterwithin*{thm}{part}
\counterwithin*{figure}{part}

\tikzset{meter/.append style={draw, inner sep=10, rectangle, font=\vphantom{A}, minimum width=30, line width=.8, path picture={\draw[black] ([shift={(.1,.3)}]path picture bounding box.south west) to[bend left=50] ([shift={(-.1,.3)}]path picture bounding box.south east);\draw[black,-latex] ([shift={(0,.1)}]path picture bounding box.south) -- ([shift={(.3,-.1)}]path picture bounding box.north);}}}
\tikzset{roundnode/.append style={circle, draw=black, fill=gray!20, thick, minimum size=10mm}}
\tikzset{squarenode/.style={rectangle, draw=black, fill=none, thick, minimum size=10mm}}

\definecolor{Blues5seq1}{RGB}{239,243,255}
\definecolor{Blues5seq2}{RGB}{189,215,231}
\definecolor{Blues5seq3}{RGB}{107,174,214}
\definecolor{Blues5seq4}{RGB}{49,130,189}
\definecolor{Blues5seq5}{RGB}{8,81,156}

\definecolor{Greens5seq1}{RGB}{237,248,233}
\definecolor{Greens5seq2}{RGB}{186,228,179}
\definecolor{Greens5seq3}{RGB}{116,196,118}
\definecolor{Greens5seq4}{RGB}{49,163,84}
\definecolor{Greens5seq5}{RGB}{0,109,44}

\definecolor{Reds5seq1}{RGB}{254,229,217}
\definecolor{Reds5seq2}{RGB}{252,174,145}
\definecolor{Reds5seq3}{RGB}{251,106,74}
\definecolor{Reds5seq4}{RGB}{222,45,38}
\definecolor{Reds5seq5}{RGB}{165,15,21}

\allowdisplaybreaks

\usepackage[most,breakable]{tcolorbox}
\makeatletter

\def\boxed@gobegin#1{\def\@tempa{#1}\def\@tempb{orange}\ifx\@tempa\@tempb\begin{tcolorbox}[colback=red!15,colframe=orange!70,breakable,enhanced]\else\begin{tcolorbox}[colback=Blues5seq1,colframe=Blues5seq5,breakable,enhanced]\fi}
\def\boxed@gobegin@empty{\begin{tcolorbox}[colback=Blues5seq1,colframe=Blues5seq5,breakable,enhanced]}
\makeatother

\begin{document}

\title{Central Limit Theorem for Bosonic Quantum Channels}

\author{Hami Mehrabi}
\email{hamimhrb@gmail.com}
\affiliation{Cornell University, Ithaca, New York 14850, USA}

\author{Ludovico Lami}
\email{ludovico.lami@gmail.com}
\affiliation{Scuola Normale Superiore, Piazza dei Cavalieri 7, 56126 Pisa, Italy}

\author{Mark M. Wilde}
\email{wilde@cornell.edu}
\affiliation{Cornell University, Ithaca, New York 14850, USA}

\begin{abstract}
In this paper, we develop an extension of the Central Limit Theorem (CLT) to the setting of bosonic quantum channels. This extension provides a deeper understanding of Gaussian bosonic channels as extremal objects. Using our CLT for bosonic quantum channels, we recover both the classical CLT and the CLT for bosonic quantum states, thereby offering a unified perspective that connects classical probability theory with continuous-variable quantum systems. Moreover, using our result, we can provide necessary uncertainty relations that every physical (possibly non-Gaussian) bosonic quantum channel must satisfy. As another application of our limit theorems, we derive tight lower bounds on the energy-constrained quantum capacity of linear bosonic channels by relating it to the capacity of their associated Gaussian bosonic channels, further reinforcing the role of Gaussian channels as extremal.
\end{abstract}

\maketitle

The Central Limit Theorem (CLT) stands as one of the most fundamental results in classical probability theory. It states that for a sequence of centered (i.e., zero-mean), independent and identically distributed (i.i.d.)\ random variables $(X_k)_{k\geq 1}$ with finite second-order moments, the normalized sum $X^{\boxplus n} \coloneqq \frac{X_1 + \cdots + X_n}{\sqrt n}$ converges to a centered Gaussian random variable $X_G$ with the same second-order moments as $X_1$~\cite{Feller}. 

Motivated by extending this fundamental result to the quantum setting, Cushen and Hudson~\cite{CH71} established a quantum version of the CLT for bosonic systems. The Cushen--Hudson quantum CLT (qCLT) asserts that for a centered $m$-mode bosonic quantum state $\rho$ with finite second-order moments, the quantum states $\rho^{\boxplus n} = \Tr_{\neg 1} \left[ U_n \rho^{\otimes n} U_n^\dagger \right]$ converge (weakly) to a centered Gaussian quantum state $\rho_G$ with the same second-order moments as $\rho$. Here $U_n$ denotes the $n$-beam splitter unitary, and $\rho^{\boxplus n}$ should be viewed as the quantum counterpart of the normalized sum of i.i.d.~random variables in the classical setting.

In this paper, we extend the Cushen--Hudson qCLT to the setting of bosonic quantum channels. We begin by using a natural construction in which $n$ copies of a bosonic quantum channel $\NN$ are sandwiched by an $n$-beam splitter unitary and its inverse as our definition of the $n$-fold symmetric convolution of $\NN$ with itself. With this construction in place, under some mild moment assumptions, we then prove that the $n$-fold symmetric convolution of $\NN$ with itself converges to a Gaussian bosonic channel, which we refer to as the Gaussification of $\NN$. 

To the best of our knowledge, the idea of sandwiching by $n$-beam splitter unitaries originates in~\cite{gisin2005error}, where it was introduced in the context of error filtration. We also point to~\cite{Wolf2007Quan} as another setting in which a Gaussian channel is obtained asymptotically from a generic bosonic channel. The authors of~\cite{Wolf2007Quan} constructed a Gaussian channel asymptotically using continuous-variable teleportation~\cite{Braunstein1998} and showed that the quantum capacity of $\NN$ is lower bounded by the capacity of the resulting Gaussian channel. Our construction is arguably preferable as the definition of the $n$-fold symmetric convolution because it relies solely on passive operations, in contrast to the teleportation-based approach of~\cite{Wolf2007Quan}. Moreover, the Gaussian channel obtained in~\cite{Wolf2007Quan} is not uniquely determined by $\NN$, and it depends heavily on the choice of entangled resource state used in the teleportation protocol.

Using our generalization of qCLT to bosonic quantum channels, we obtain a unified framework that recovers both the classical CLT in probability theory and its counterpart in continuous-variable quantum systems, highlighting its role as a fundamental limit theorem. On the side of applications, our limit theorem allows for formulating necessary uncertainty relations that any \emph{physical} bosonic quantum channel must satisfy. Moreover, it enables the derivation of entropic inequalities involving linear bosonic channels and their Gaussifications. For instance, we show that the energy-constrained quantum capacity of a linear bosonic channel is lower bounded by that of its Gaussification. Our method, however, applies similarly to other entropic quantities, such as the energy-constrained classical capacity.

Moreover, the definition of symmetric convolutions for channels is useful beyond limit theorems. For example, by using the two-fold symmetric convolution, we can generalize the classical Kac--Bernstein theorem~\cite{Kac1939OnAC, bernstein1941property} to the setting of bosonic quantum channels. The classical Kac--Bernstein theorem states that if two independent random variables $X$ and $Y$ are such that $X+Y$ and $X-Y$ are also independent, then $X$ and $Y$ must be Gaussian random variables. Using our definition of the symmetric convolution of bosonic quantum channels, we can obtain an analogous characterization in the setting of bosonic quantum channels: If the extension of the two-fold symmetric convolution of a channel $\NN$ maps product states to product states, then $\NN$ must be a Gaussian channel.

To gain some intuition about our construction and main result, it is helpful to examine how the symmetric convolution acts on an additive-noise channel in the Heisenberg picture. An additive-noise channel transforms an annihilation operator as $ \ac \to \ac + W$, where $W$ is a random variable describing the additive noise. By preprocessing $n$ input modes with the $n$-beam splitter, applying $n$ i.i.d.~uses of the additive-noise channel, and then acting with the inverse of the $n$-beam splitter while tracing out all modes except the first, the resulting Heisenberg-picture transformation on the first mode is $ \ac \to \ac + W^{\boxplus n}$. Thus the procedure converts an additive-noise channel with noise $W$ into a new additive-noise channel with noise $W^{\boxplus n}$, which is known from the classical CLT to converge to a Gaussian distribution. While this brief discussion is intended only to provide intuition, our approach to a quantum CLT for channels is fully rigorous and is carried out using the Weyl representation of bosonic systems.

In the following, after reviewing the basics of bosonic quantum systems, we define the symmetric convolution method for bosonic quantum channels. We then present our main result, the qCLT for bosonic quantum channels, together with illustrative examples that clarify the construction and demonstrate how both the classical CLT and the Cushen--Hudson qCLT can be recovered within our framework. After that, we turn to applications, including the uncertainty relations for bosonic quantum channels. In the final part, we apply our limit theorem to derive lower bounds on the quantum capacity of linear bosonic channels. The proof of our qCLT for bosonic quantum channels is deferred to the appendix.

\section*{Preliminaries}

In this paper, we primarily work with single-mode bosonic quantum states, and for clarity we state all results in the single-mode setting. However, none of our arguments rely on this restriction, and all results extend directly to the multimode case. The Hilbert space of a single-mode bosonic quantum system is isomorphic to the space of all complex-valued functions over $\R$ with finite second norm. We use the notation $\HH \cong L^2(\R)$ to denote this Hilbert space. The \emph{quadrature operators} for a single mode are denoted by the vector $\mathbf{R} \coloneqq  \left( \x, \p\right)^\top$, and the associated \emph{annihilation} and \emph{creation} operator are defined as $$\ac \coloneqq \frac{1}{\sqrt 2} \left( \x + i \p\right), \quad \ac^\dagger \coloneqq \frac{1}{\sqrt 2} \left( \x - i \p\right).$$
These operators satisfy the \emph{canonical commutation relations (CCR)}, which can be written in coordinate form as $[\mathbf R , \mathbf R^\top] = i \Omega$, where $\Omega$ is the symplectic form given by
\bb
\Omega \coloneqq \begin{pmatrix}
	  0 & 1\\
	  -1 & 0 
	  \end{pmatrix}.
\ee
The eigenstates of the photon-number operator $\mathbf{N} \coloneqq \ac^\dagger \ac$ consist of \emph{Fock states}. For each non-negative integer $\ell$, the Fock state corresponding to a state with $\ell$ photons is denoted by $\ket{\ell}$, and the state $\ket{0}$ is called \emph{vacuum state}. The action of the annihilation and creation operators on Fock states is given by 
\bb
\ac \ket{\ell} = \sqrt{\ell} \ket{\ell-1}, \quad \ac^\dagger \ket{\ell} = \sqrt{\ell+1} \ket{\ell+1}.
\ee

We use $\T(\HH)$ to denote the Banach space of trace-class operator acting on $\HH$, while its dual, the Banach space of bounded operators is denoted by $\BC(\HH)$ \cite{Hall-Book}. A single mode bosonic quantum state is a positive semi-definite trace-class operator acting on $\HH$ with trace equal to $1$. 

For a given bosonic quantum state $\rho$, the vector of its first-order moments is defined as $\dd(\rho) \coloneq \Tr \!\left[ \rho \mathbf R\right]$, and its covariance matrix is given by
\bb
	\gamma(\rho) \coloneqq \Tr\left[\rho \{ \mathbf R - \dd(\rho), \mathbf R^\top - \dd(\rho)^\top\}\right],
\ee
where $\{.,.\}$ is the coordinate-wise anti-commutator. The bosonic quantum state $\rho$ is called \emph{centered} whenever the vector of its first-order moments is the zero vector. In general, for every $\kappa \geq 0$, we define the $\kappa$-order moment of $\rho$ as $M_\kappa(\rho) \coloneqq \Tr\left[ \rho \mathbf{N}^{\kappa/2}\right]$. It is straightforward to verify that as long as $\rho$ has finite second-order moment, its covariance matrix $\gamma(\rho)$ is well-defined. The \emph{Robertson--Schr\"{o}dinger uncertainty relation} then implies that $\gamma(\rho) + i \Omega \geq 0$ for every bosonic quantum state $\rho$. Also, assuming finiteness of the second-order moment of $\rho$, we can define the \emph{von Neumann entropy} of $\rho$ as $S(\rho) \coloneqq -\Tr \left[ \rho \log \rho\right]$.

For $z \in \C$, the displacement operator $D_z$ is defined as $D_z \coloneqq  \exp \!\left( z \ac^\dagger - \bar{z} \ac \right)$. By the action of the displacement operator $D_\alpha$ on the vacuum state, we define coherent states as $\ket{\alpha} \coloneqq D_\alpha \ket{0}$. Also, using these displacement operators, the characteristic function of a trace-class operator $T$ is defined as $$\chi_T(z) \coloneqq \Tr\left[T D_z\right].$$

A bosonic quantum state is called Gaussian whenever its characteristic function is a Gaussian function, meaning it can be written as the exponential of a second-order polynomial. A Gaussian quantum state is uniquely determined by its first-order moments and its covariance matrix. The characteristic function of a Gaussian bosonic quantum state $\sigma$ takes the form
\bb\label{eq:CharGaussian}
	\chi_{\sigma}(z) = \exp\left(-\frac 14 \hat{z}^\dagger \Lambda^\dagger \boldsymbol{\gamma}(\sigma) \Lambda  \hat{z} + i\dd(\sigma)^\top \Lambda \hat z\right),
\ee
where $\hat{z} = (z, \bar{z})^\top$ and $\Lambda = \frac{1}{\sqrt 2} \begin{pmatrix}
	  -i & i\\
	  -1 & -1 
	  \end{pmatrix}.$

A bosonic quantum channel $\NN$ is a linear map from $\T(\HH)$ to $\T(\HH)$ that is completely positive and trace preserving. Keeping in mind that $\BC(\HH)$ is the dual space of $\T(\HH)$, we define $\NN^\dagger$ to be the adjoint of $\NN$, namely the linear map from $\BC(\HH)$ to $\BC(\HH)$ satisfying
\[
\Tr\left[T \NN^\dagger(B)\right] = \Tr\left[\NN(T) B\right],
\]
for all $T \in \T(\HH)$ and $B \in \BC(\HH)$. Also, we call a bosonic quantum channel $\NN$ \emph{centered} if $\NN(\ketbra{0})$ is centered as a bosonic quantum state.

A quantum channel $\NN$ is called Gaussian if it maps Gaussian quantum states to Gaussian quantum states. As with Gaussian quantum states, a Gaussian channel $\NN_G$ is uniquely determined by its action on first-order moments and the covariance matrix. For a centered single-mode Gaussian channel, this action is completely characterized by real $2 \times 2$ matrices $X$ and $Y$, with $Y=Y^\top$, which specify how the channel transforms covariance matrices and the vector of first-order moments via
\bb\label{eq:GauusianChannel_Covariance}
	\gamma( \NN_G( \rho)) = X \gamma(  \rho) X^\top + Y, \quad \dd( \NN_G( \rho)) = X \dd(\rho).
\ee
A bosonic Gaussian channel is physical if and only if the uncertainty relation $Y + i  \Omega  \geq i X \Omega X^\top $ holds. Note that this latter condition implies that $Y\geq 0$, as one can swiftly see by taking its real part.

\section*{Channel Symmetric Convolution} 

\begin{figure}
    \centering
    \includegraphics[width=0.90\linewidth]{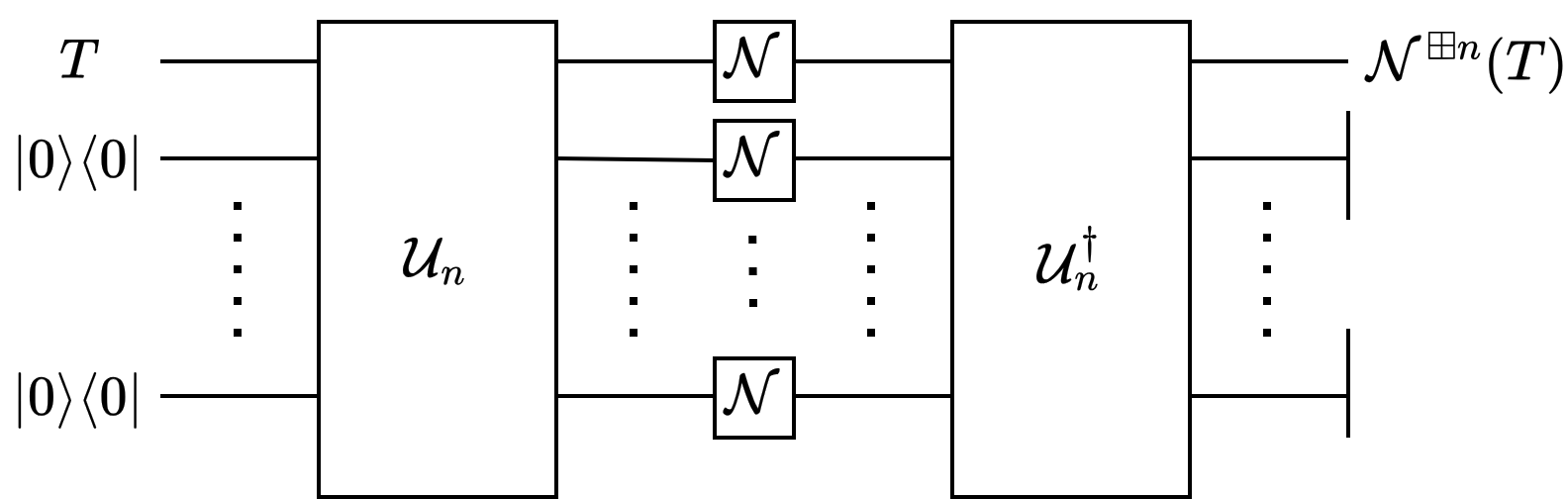}
    \caption{Symmetric convolution of channel $\NN$ with itself.}
    \label{fig:SymmetricConvolution}
\end{figure}

As mentioned earlier, we define the $n$-fold symmetric convolution of a channel $\NN$ with itself by exploiting the $n$-beam splitter unitary $U_n$. This unitary acts on an $n$-fold tensor-product system, where each subsystem is isomorphic to a single-mode bosonic quantum system. In the Heisenberg picture, for every $\vec{\mathbf{z}} = (z_1, \ldots, z_n)^\top \in (\C)^n$, the $n$-beam splitter unitary $U_n$ acts on displacement operators as
\bb
U_n \left( D_{z_1} \otimes \cdots \otimes D_{z_n} \right) U_n^\dag = D_{V_n \vec{\mathbf{z}}}, \quad (V_n)_{jk} \coloneqq \frac{e^{\frac{2 \pi i}{n}(j-1)(k-1)}}{\sqrt{n}}.
\label{eq:def-U-n}
\ee
Using this unitary, we define the $n$-fold symmetric convolution of a single-mode bosonic quantum channel $\NN$ with itself as the channel acting on $T \in \T(\HH)$ by
\bb\label{def:ConvChannels}
\NN^{\,\boxplus n}(T) \coloneqq \Tr_{\neg 1} \! \left[U_n \NN^{\otimes n} \!\left(U_n^\dag \left( T \otimes \ketbra{0}^{\otimes (n-1)} \right) U_n \right) U_n^\dag \right]\! ,
\ee
where $\ketbra{0}$ denotes the single-mode vacuum state, and $\Tr_{\neg 1}$ denotes the partial trace over all subsystems except the first; see Fig.~\ref{fig:SymmetricConvolution} for visual depiction of the transformation.

Our first observation is that if $\NN_G$ is a (centered) Gaussian channel, then $\NN_G^{\boxplus n}= \NN_G$ for every $n \geq 1$. To see this, we begin with a broader class of bosonic Gaussian channels, namely, linear bosonic channels. A bosonic quantum channel $\NN$ is called linear if its adjoint acts on displacement operators as
\bb\label{eq:AdjointLinearChannel}
	\NN^\dagger(D_z) = f(z) D_{L(z)},
\ee
where $f(z) \equiv f(\hat{z})$, for $\hat{z} \equiv (z, \bar{z})^\top$, is called the scaling function associated with $\NN$ and $L(z) \equiv L(\hat z)$ is a linear transformation~(see, e.g.,~\cite{Holevo1972QuantumCommunicationChannels, PhysRevA.63.032312}). For necessary and sufficient conditions on $f$ and $L$ that ensure $\NN$ is a legitimate bosonic quantum channel, we refer the reader to~\cite[Lemma 3]{lami2018all}. It is straightforward to see that all Gaussian channels are linear. Moreover, the centered Gaussian ones are precisely those whose adjoint action satisfies $\NN_G^\dagger(D_z) = f_G(z) D_{L(z)}$, where $f_G$ is a \emph{centered} Gaussian function, meaning that it is the exponential of a polynomial consisting only of second-order terms.

Now, for a general linear bosonic channel $\NN$ with adjoint action given by~\eqref{eq:AdjointLinearChannel}, we can express the action of the $n$-fold symmetric convolution of $\NN$ with itself at the level of characteristic functions as
\bb\label{eq:SymmetricConvolutionLinear}
    &\chi_{\NN^{\boxplus n}(T)}\left(z\right) = \chi_{U_n \NN^{\otimes n} \left( U_n^\dagger \left(T \otimes  \ketbra{0}^{\otimes (n-1)} \right) U_n\right) U_n^\dagger}\left(z , 0 , \cdots, 0\right) \\
    & = \chi_{\NN^{\otimes n} \left( U_n^\dagger \left(T \otimes \ketbra{0}^{\otimes (n-1)} \right) U_n\right)}\left(\frac{z}{\sqrt n}, \cdots , \frac{z}{\sqrt n}\right) \\
    & = f\!\left(\frac{z}{\sqrt n}\right)^n \cdot \chi_{ U_n^\dagger \left(T \otimes \ketbra{0}^{\otimes (n-1)} \right) U_n} \left(\frac{L(z)}{\sqrt n}, \cdots , \frac{L(z)}{\sqrt n}\right) \\
    & = f\!\left(\frac{z}{\sqrt n}\right)^n \cdot \chi_{ T \otimes \ketbra{0}^{\otimes (n-1)} } \left(
    L(z), 0 , \cdots , 0\right) \\
    & = f\!\left(\frac{z}{\sqrt n}\right)^n \cdot \chi_T\left(L(z)\right).
\ee
Thus, for a linear bosonic channel $\NN$ with scaling function $f$ and linear transformation $L$, the channel $\NN^{\boxplus n}$ is itself a linear bosonic channel with the same linear transformation $L$, while its scaling function is modified according to $f(z) \to f(z/\sqrt{n})^n$. We also note from the expression in the penultimate line of~\eqref{eq:SymmetricConvolutionLinear} that having the vacuum in the other $n-1$ ports is not necessary, and one may place any other unit-trace operator there instead of the vacuum state. We refer to this feature as the no-signalling property of the symmetric convolution for linear bosonic channels.

From the above behavior of linear bosonic channels, we can immediately conclude that if $\NN_G$ is a centered Gaussian channel, then $\NN_G^{\boxplus n} = \NN_G$. This follows from the fact that $f_G$ is a centered Gaussian function, and such functions satisfy $f_G(z/\sqrt{n})^n = f_G(z)$.

This is a natural property to expect from any reasonable symmetric convolution: it should not alter a centered Gaussian channel. For comparison, the same phenomenon appears in both the classical and quantum settings. In the classical case, if $(X_k)_{k\geq 1}$ is an i.i.d.~sequence of centered Gaussian random variables, then $X^{\boxplus n}$ has the same centered Gaussian distribution for every $n \geq 1$. Likewise, for quantum states, if $\sigma$ is a centered single-mode Gaussian quantum state, then $\sigma^{\boxplus n} = \sigma$ for every $n \geq 1$. In contrast, the procedure proposed in~\cite{Wolf2007Quan} preserves Gaussian channels only in the unphysical limit where the entangled resource state has infinite squeezing.

As another example, consider the replacement channel $\NN_\rho(T) = \Tr[T]\, \rho$, which replaces every unit-trace input with the fixed bosonic quantum state $\rho$. For this channel, we have
\bb
&\NN^{\,\boxplus n}(T) = \Tr_{\neg 1} \! \left[U_n \NN^{\otimes n} \!\left(U_n^\dag\, \left( T \otimes \ketbra{0}^{\otimes (n-1)} \right)\, U_n \right) U_n^\dag \right]\\
&= \Tr[T]  \Tr_{\neg 1} \! \left[U_n \rho^{\otimes n} U_n^\dag \right] = \Tr[T] \, \rho^{\boxplus n}.
\ee
The above equation shows that $\NN^{\boxplus n}$ is also a replacement channel, one that replaces every unit-trace input with $\rho^{\boxplus n}$. This shows that our notion of convolution for channels reduces to the one for states when applied to the special class of replacer channels.

For the last example, suppose that $\NN_W$ is an additive-noise channel, where $W$ is $\C$-valued classical random variable, and the channel $\NN_W$ adds this classical noise to the input as
\bb
	\NN_W (T) \coloneqq \int \dd^{2} \omega\ p_W(\omega) D_\omega T D_\omega^\dagger.
\ee
At the level of characteristic functions, we have 
\bb\label{eq:CharAdditiveNoise}
\chi_{\NN_W (T)}(z) = \chi_T(z) \, \chi_W\big(z\big),
\ee
where $\chi_W\big(z\big) \coloneqq \int p_W(\omega) e^{\sqrt{t}(z^\top \bar \omega - \bar z^\top \omega)} \dd^{2}\omega$ is the symplectic characteristic function of $W$. We see that additive-noise channels also belong to the class of linear bosonic channels. Moreover, applying~\eqref{eq:SymmetricConvolutionLinear}, we find that $\NN_W^{\boxplus n}$ is again a linear bosonic channel, with scaling function $\chi_W(z/\sqrt{n})^n$ which is precisely the symplectic characteristic function of the $n$-fold symmetric summation of $W$ with itself. Thus, $\NN_W^{\boxplus n}$ is again an additive-noise channel, now with noise $W^{\boxplus n}$.

\section*{CLT for Bosonic Quantum Channels}

In this section, we state our main result regarding the qCLT for bosonic quantum channels. This theorem is in the same vein as the classical CLT and the Cushen--Hudson qCLT for bosonic quantum states. As in those settings, the qCLT for bosonic quantum channels asserts that the $n$-fold symmetric convolution of a channel $\NN$ with itself, as defined earlier, converges to a Gaussian bosonic channel $\NN_G$, which we refer to as the \emph{Gaussification} of $\NN$. Analogous to the classical and Cushen--Hudson limit theorems, a suitable moment assumption is required to guarantee this convergence. The following theorem precisely states our qCLT for bosonic quantum channels.

\begin{thm}\label{Thm:qCLTChannel}
Let $\NN$ be a centered single-mode bosonic quantum channel such that the state $\NN\left(\ketbra{0}\right)$ has a finite second-order moment, and $\NN\left(\ketbraa{1}{0}\right)$ has a finite first-order moment. Then $\NN^{\boxplus n}$ converges, in the strong topology, to a Gaussian bosonic channel $\NN_G$. That is, for every single-mode bosonic quantum state $\rho$, $$\lim_{n \to \infty} \NN^{\boxplus n} (\rho) \to \NN_G(\rho), \quad \text{in trace norm.}$$ Moreover, the corresponding matrices $X$ and $Y$ associated with $\NN_G$ are given by
\bb\label{eq:XqCLT}
X &= \frac{1}{\sqrt2} \Tr \left[ \mathbf R \begin{pmatrix}  \NN\left(\ketbraa{0}{1} + \ketbraa{1}{0}\right) \\[4pt] \NN\left(-i\ketbraa{0}{1} + i\ketbraa{1}{0}\right) \end{pmatrix}^\top \right], \\
Y &= V - XX^\top,
\ee
where $\mathbf{R} = \left( \x, \p\right)^\top$ is the vector of quadrature operators and $V$ is the covariance matrix of $\NN(\ketbra{0})$. 
\end{thm}

As we mentioned earlier, Theorem~\ref{Thm:qCLTChannel} generalizes both the classical CLT and the quantum CLT for bosonic quantum states. To see how it recovers the classical CLT, let $W$ be a centered $\C$-valued random variable with finite second-order moments, and let $\NN_W$ be the associated additive-noise channel with noise $W$. As noted earlier, using~\eqref{eq:SymmetricConvolutionLinear} together with~\eqref{eq:CharAdditiveNoise}, we can directly verify that $\NN_W^{\boxplus n}$ is the additive-noise channel with noise $W^{\boxplus n}$. 

Moreover, it is straightforward to check that for the channel $\NN_W$, the parameters of the Gaussification obtained from~\eqref{eq:XqCLT} coincide with those of the Gaussian additive-noise channel $\NN_{W_G}$, where $W_G$ is the centered complex Gaussian random variable with the same covariance matrix as $W$. Therefore, Theorem~\ref{Thm:qCLTChannel} implies that $W^{\boxplus n}$ converges to $W_G$, recovering the classical CLT.

To see how the Cushen--Hudson qCLT for bosonic quantum states follows from Theorem~\ref{Thm:qCLTChannel}, let $\rho$ be a centered single-mode bosonic quantum state with finite second-order moments, and let $\NN_\rho$ be the replacement channel with output $\rho$ for unit-trace inputs. As noted earlier, $\NN_\rho^{\boxplus n}$ is again a replacement channel, now with output $\rho^{\boxplus n}$. Since $\rho$ is centered and has finite second-order moments, all assumptions of Theorem~\ref{Thm:qCLTChannel} are satisfied. Applying Theorem~\ref{Thm:qCLTChannel}, we conclude that the sequence of replacement channels with outputs $\rho^{\boxplus n}$ converges to a Gaussian channel whose parameters are determined by~\eqref{eq:XqCLT} applied to $\NN_\rho$. A direct computation shows that the resulting matrix $X$ vanishes, while $Y$ is precisely the covariance matrix of $\rho$. Therefore, the Gaussification of $\NN_\rho$ is the replacement channel with output $\rho_G$, the Gaussification of $\rho$. Consequently, Theorem~\ref{Thm:qCLTChannel} implies that $\rho^{\boxplus n}$ converges to $\rho_G$ in trace distance, which is exactly the Cushen--Hudson quantum CLT for bosonic quantum states.

Theorem~\ref{Thm:qCLTChannel} can also be used to derive a necessary uncertainty relation that any physical completely positive linear map $\NN\colon \T(\HH) \to \T(\HH)$ must satisfy. The following corollary follows directly from Theorem~\ref{Thm:qCLTChannel} together with the uncertainty relations for Gaussian channels.

\begin{cor}
Let $\NN$ be a centered single-mode bosonic quantum channel satisfying the assumptions of Theorem~\ref{Thm:qCLTChannel}. Then a necessary condition that $\NN$ must satisfy is
$Y + i \Omega \geq i X \Omega X^\top $, where $X$ and $Y$ are the matrices associated with the Gaussification of $\NN$, as defined in~\eqref{eq:XqCLT}.
\end{cor}

\section*{Energy-Constrained Quantum Capacity}
In this section, we present an application of the quantum CLT for bosonic quantum channels stated in Theorem~\ref{Thm:qCLTChannel}, namely, a lower bound on the energy-constrained quantum capacity of linear bosonic channels. Starting from the definition of energy-constrained coherent information, for a single-mode bosonic quantum channel $\NN^{A \to B}$, the energy-constrained coherent information at energy $E \geq 0$ is defined as
\bb
\tilde Q\left(\NN, E\right) \coloneqq \sup I\left(R \rangle B\right)_{\left( \Id_R \otimes \NN_{A \to B} \right)\varphi_{RA}} ,
\ee
where the supremum is taken over all pure bosonic quantum states $\varphi_{RA}$ satisfying $\Tr \left[ \varphi_{RA} \mathbf{N}_A\right] \leq E$, with $\mathbf{N}_A$ the photon-number operator on system $A$~\cite{giovannetti2003broadband}. The quantity $I\left(R \rangle B\right)_{\omega_{RB}} \coloneqq S(\omega_B) - S(\omega_{RB})$ denotes the coherent information. The energy-constrained quantum capacity at energy $E \geq 0$ is then obtained by regularizing the coherent information:
\bb
C_Q \left(\NN, E\right) \coloneqq \lim_{n\to \infty} \frac{1}{n} \tilde Q\left(\NN^{\otimes n}, n \times E\right).
\ee

The following corollary states the lower bound on the energy-constrained quantum capacity that follows from Theorem~\ref{Thm:qCLTChannel}.

\begin{cor}\label{Cor:LowerBoundCapacity}
Let $\NN$ be a single-mode linear bosonic channel with linear transformation $L$ and scaling function $f$ as defined in~\eqref{eq:AdjointLinearChannel}. Suppose $\NN$ satisfy the assumptions in Theorem~\ref{Thm:qCLTChannel}, and $f$ is an even function such that $f(z) = f(-z)$. Then, for every $E \geq 0$, we have
\bb
	C_Q \left(\NN, E\right) \geq C_Q \left(\NN_G, E\right),
\ee
where $\NN_G$ is the Gaussification of $\NN$.
\end{cor}

\begin{figure}
    \centering
    \includegraphics[width=0.9\linewidth]{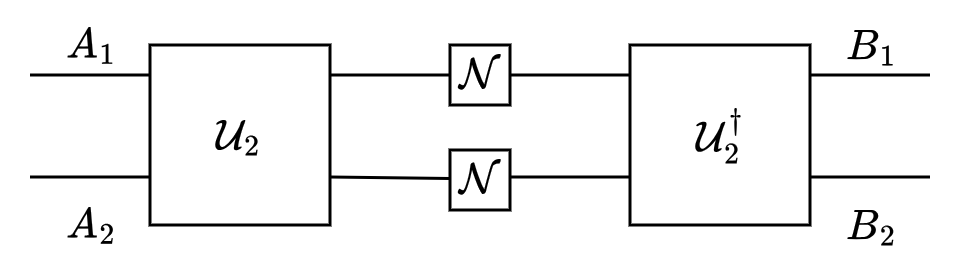}
    \caption{Extension of the $2$-fold symmetric convolution.}
    \label{fig:Linear2Symmetric}
\end{figure}

To illustrate this, consider the extension $\tilde \NN^{A_1A_2 \to B_1B_2} \coloneqq  \T(\HH^{\otimes 2}) \to \T(\HH^{\otimes 2})$ to be the quantum channel $\tilde \NN^{A_1A_2 \to B_1B_2}(T_{A_1A_2}) \coloneqq  U_2 \NN^{\otimes 2} \left( U_2^\dagger \left(T_{A_1A_2}\right) U_2\right) U_2^\dagger$; See Fig.~\ref{fig:Linear2Symmetric} for a visual depiction. 

Using the expression in the penultimate line of~\eqref{eq:SymmetricConvolutionLinear} for linear bosonic channels, we obtain the no-signalling property that the channel $\tilde \NN^{A_1 \to B_1}$ is independent of the input on $A_2$. By definition, the reduced channel $\tilde \NN^{A_1 \to B_1} = \NN^{\boxplus 2}$. 

It is also straightforward to see that the same property holds for the channel $\tilde \NN^{A_2 \to B_2}$ that there is no signaling from $A_1$ to $B_2$. Moreover, under the additional assumption that $f(z)= f(-z)$, the channel $\tilde \NN^{A_2 \to B_2}$ is identical to $\NN^{\boxplus 2}$. 

Now using the superadditivity of coherent information, we have
\begin{equation}
    I(R_1 R_2 \rangle B_1 B_2 ) \geq I(R_1 \rangle B_1) + I(R_2\rangle B_2),
\end{equation}
where here $R_1$ and $R_2$ are purifying systems for $A_1$ and $A_2$~\cite{datta2012quantum}. Taking the energy-constrained supremum over all input states, and using the facts that the channels $A_1 \to B_1$ and $A_2 \to B_2$ are both identical to $\NN^{\boxplus 2}$, and that, by the no-signalling property, the channel $A_1 \to B_1$ is independent of the input on $A_2$ (and similarly $A_2 \to B_2$ is independent of the input on $A_1$), we obtain
\begin{equation}
    \tilde Q(\tilde\NN^{A_1A_2 \to B_1B_2}, 2\times E) \geq 2 \tilde Q(\NN^{\boxplus 2}, E) 
\end{equation}
for the channel coherent information $Q$ and every $E >0$. Removing unitaries in the definition of $\tilde\NN^{A_1A_2 \to B_1B_2}$, we obtain $\tilde Q( \NN^{\otimes 2},2E ) \geq 2 \tilde Q(\NN^{\boxplus 2}, E)$.
Now by induction and the fact that $\NN^{\boxplus 2^{k+1}} = (\NN^{\boxplus 2^k})^{\boxplus 2}$~\footnote{For every pair of positive integers $n_1, n_2$, and every coherent state $\ketbra{\alpha}$, using~\eqref{eq:SymmetricConvolutionCoherent} we have $\NN^{\boxplus (n_1 \times n_2)}(\ketbra{\alpha}) = \NN(\ketbra{\alpha/\sqrt{n_1 \times n_2}})^{\boxplus (n_1 \times n_2)}$. Also, we can write
\bb
&\left(\NN^{\boxplus n_1}\right)^{\boxplus n_2} (\ketbra{\alpha}) = \left(\NN^{\boxplus n_1} (\ketbra{\alpha/\sqrt{n_2}})\right)^{\boxplus n_2}\\
& = \left(\left(\NN (\ketbra{\alpha/\sqrt{n_1 \times n_2}})\right)^{\boxplus n_1}\right)^{\boxplus n_2} \\
&= \NN(\ketbra{\alpha/\sqrt{n_1 \times n_2}})^{\boxplus (n_1 \times n_2)}.
\ee
Following the fact that the span of coherent states are dense in the Banach space of trace class operators, we can conclude that $\NN^{\boxplus (n_1 \times n_2)}= \left(\NN^{\boxplus n_1}\right)^{\boxplus n_2}$}, we get 
\begin{equation}
    \frac{\tilde Q( \NN^{\otimes 2^k}, 2^k \times E)}{2^k} \geq \tilde Q(\NN^{\boxplus 2^k}, E),
\end{equation}
for every $k \geq 1$. As $k \to \infty$, we get 
\begin{equation}
    C_Q(\NN, E) = \lim_{k\to \infty} \frac{\tilde Q(\NN^{\otimes k}, k E)}{k} \geq \tilde Q(\NN_G, E),
\end{equation}
where $C_Q$ denotes the quantum capacity. Note that, for the right-hand side, we use continuity bounds for energy-constrained channel coherent information with respect to the energy-constrained diamond norm~\cite{winter2017energy}. Thus, using the fact that $C_Q(\NN, E)= \frac 1k C_Q(\NN^{\otimes k}, k\times E)$ for every $k\geq 1$ and $(\NN^{\otimes k})_G = \NN_G^{\otimes k}$, we have 
\begin{align}
    C_Q(\NN, E) &= \lim_{k \to \infty } \frac{C_Q(\NN^{\otimes k}, k\times E)}{k}\\
    &\geq \lim_{k \to \infty } \frac{\tilde Q(\NN_G^{\otimes k}, k\times E)}{k} = C_Q(\NN_G, E).
\end{align}

The lower bound on the energy-constrained quantum capacity in Corollary~\ref{Cor:LowerBoundCapacity} is in the same spirit as the capacity result of~\cite{Wolf2007Quan}. In addition to this parallel, the lower bound provided in Corollary~\ref{Cor:LowerBoundCapacity} concerns the \emph{energy-constrained} quantum capacity, which is more relevant for realistic physical implementations than unconstrained achievable rates. However, the linearity of the channel $\NN$ is crucial in Corollary~\ref{Cor:LowerBoundCapacity}. In fact, the lower bound in Corollary~\ref{Cor:LowerBoundCapacity} may fail for certain non-linear bosonic channels. For example, let $\NN_p \colon\T(\HH) \to \T(\HH)$ be the bosonic dephasing channel associated with a classical probability distribution $p$, defined as
\[
\NN_p(T)\coloneqq \int_{0}^{2\pi} e^{i \theta \ac^\dagger \ac }T e^{-i \theta \ac^\dagger \ac} p(\theta) \dd \theta.
\]
We can write the action of $\NN_p$ on Fock states as $\NN_p(\ketbraa{n}{m}) = \Phi_p(n-m) \ketbraa{n}{m}$, where $\Phi_p(k) \coloneqq \int_{0}^{2\pi} e^{i \theta k} p(\theta) \dd \theta$. Using this, we can conclude that the Gaussification of $\NN_p$ is the pure-loss channel with transmissivity parameter $\sqrt{\lambda} = \left| \Phi_p(1) \right|$. 

The quantum capacity of $\NN_p$ is $\log(2\pi) - h(p)$, where $h(p) \coloneqq -\int_0^{2\pi} p(\theta) \log(p(\theta)) \dd \theta$ is the differential entropy~\cite{lami2023exact}. On the other hand, the quantum capacity of a pure-loss channel with transmissivity $\lambda$ is $\max \{ 0, \log (\frac{\lambda}{1-\lambda})\}$~\cite{Wolf2007Quan}. Comparing the quantum capacity of the bosonic dephasing channel with that of its Gaussification, the corresponding pure-loss channel, we can choose classical probability distributions $p$ for which the quantum capacity of $\NN_p$ is strictly smaller than the quantum capacity of its Gaussification, showing that the lower bound in Corollary~\ref{Cor:LowerBoundCapacity} does not hold in general.

\medskip
\begin{acknowledgments}
We thank Zixin Huang and Hemant Mishra for helpful discussions. HM and MMW acknowledge the School of Electrical and Computer Engineering at Cornell University for support. LL acknowledges financial support from the European Union under the European Research Council (ERC Grant Agreement No.~101165230).
\end{acknowledgments}

\bibliographystyle{unsrturl}
\bibliography{CLTBIB}

\begin{thebibliography}{10}

\bibitem{Feller}
William Feller.
\newblock {\em An introduction to probability theory and its applications, volume 2}, volume~81.
\newblock John Wiley \& Sons, 1991.

\bibitem{CH71}
Clive~D. Cushen and Robin~L. Hudson.
\newblock A quantum-mechanical central limit theorem.
\newblock {\em Journal of Applied Probability}, 8(3):454--469, 1971.
\newblock \href {https://doi.org/10.2307/3212170} {\path{doi:10.2307/3212170}}.

\bibitem{gisin2005error}
Nicolas Gisin, Noah Linden, Serge Massar, and Sandu Popescu.
\newblock Error filtration and entanglement purification for quantum communication.
\newblock {\em Physical Review A}, 72(1):012338, 2005.
\newblock \href {https://doi.org/10.1103/PhysRevA.72.012338} {\path{doi:10.1103/PhysRevA.72.012338}}.

\bibitem{Wolf2007Quan}
Michael~M. Wolf, David P\'erez-Garc\'{\i}a, and Geza Giedke.
\newblock Quantum capacities of bosonic channels.
\newblock {\em Physical Review Letters}, 98:130501, March 2007.
\newblock \href {https://doi.org/10.1103/PhysRevLett.98.130501} {\path{doi:10.1103/PhysRevLett.98.130501}}.

\bibitem{Braunstein1998}
Samuel~L. Braunstein and H.~Jeffrey Kimble.
\newblock Teleportation of continuous quantum variables.
\newblock {\em Physical Review Letters}, 80:869--872, 1998.
\newblock \href {https://doi.org/10.1103/PhysRevLett.80.869} {\path{doi:10.1103/PhysRevLett.80.869}}.

\bibitem{Kac1939OnAC}
Mark Kac.
\newblock On a characterization of the normal distribution.
\newblock {\em American Journal of Mathematics}, 61:726, 1939.
\newblock \href {https://doi.org/10.2307/2371328} {\path{doi:10.2307/2371328}}.

\bibitem{bernstein1941property}
Sergei~N. Bernstein.
\newblock On a property which characterizes a {G}aussian distribution.
\newblock {\em Proceedings of the Leningrad Polytechnic Institute}, 217(3):21--22, 1941.

\bibitem{Hall-Book}
Brian~C. Hall.
\newblock {\em Quantum Theory for Mathematicians}.
\newblock Springer New York, NY, 2013.
\newblock \href {https://doi.org/10.1007/978-1-4614-7116-5} {\path{doi:10.1007/978-1-4614-7116-5}}.

\bibitem{Holevo1972QuantumCommunicationChannels}
Alexander~S. Holevo.
\newblock On the mathematical theory of quantum communication channels.
\newblock {\em Problems of Information Transmission}, 8(1):47--54, 1972.

\bibitem{PhysRevA.63.032312}
Alexander~S. Holevo and Reinhard~F. Werner.
\newblock Evaluating capacities of bosonic {G}aussian channels.
\newblock {\em Physical Review A}, 63:032312, February 2001.
\newblock \href {https://doi.org/10.1103/PhysRevA.63.032312} {\path{doi:10.1103/PhysRevA.63.032312}}.

\bibitem{lami2018all}
Ludovico Lami, Krishna~Kumar Sabapathy, and Andreas Winter.
\newblock All phase-space linear bosonic channels are approximately {G}aussian dilatable.
\newblock {\em New Journal of Physics}, 20(11):113012, 2018.
\newblock \href {https://doi.org/10.1088/1367-2630/aae738} {\path{doi:10.1088/1367-2630/aae738}}.

\bibitem{giovannetti2003broadband}
Vittorio Giovannetti, Seth Lloyd, Lorenzo Maccone, and Peter~W. Shor.
\newblock Broadband channel capacities.
\newblock {\em Physical Review A}, 68(6):062323, 2003.
\newblock \href {https://doi.org/10.1103/PhysRevA.68.062323} {\path{doi:10.1103/PhysRevA.68.062323}}.

\bibitem{datta2012quantum}
Nilanjana Datta, Min-Hsiu Hsieh, and Mark~M Wilde.
\newblock Quantum rate distortion, reverse {S}hannon theorems, and source-channel separation.
\newblock {\em IEEE Transactions on Information Theory}, 59(1):615--630, 2012.
\newblock \href {https://doi.org/10.1109/TIT.2012.2215575} {\path{doi:10.1109/TIT.2012.2215575}}.

\bibitem{Note1}
For every pair of positive integers $n_1, n_2$, and every coherent state $\mathinner {|{\alpha }\rangle }\protect \!\mathinner {\langle {\alpha }|}$, using~\protect \eqref {eq:SymmetricConvolutionCoherent} we have $\protect \pazocal {N}^{\boxplus (n_1 \times n_2)}(\mathinner {|{\alpha }\rangle }\protect \!\mathinner {\langle {\alpha }|}) = \protect \pazocal {N}(\mathinner {|{\alpha /\protect \sqrt {n_1 \times n_2}}\rangle }\protect \!\mathinner {\langle {\alpha /\protect \sqrt {n_1 \times n_2}}|})^{\boxplus (n_1 \times n_2)}$. Also, we can write \begin {equation}\begin {aligned}\protect \hspace {0pt}&\left (\protect \pazocal {N}^{\boxplus n_1}\right )^{\boxplus n_2} (\mathinner {|{\alpha }\rangle }\protect \!\mathinner {\langle {\alpha }|}) = \left (\protect \pazocal {N}^{\boxplus n_1} (\mathinner {|{\alpha /\protect \sqrt {n_2}}\rangle }\protect \!\mathinner {\langle {\alpha /\protect \sqrt {n_2}}|})\right )^{\boxplus n_2}\\ & = \left (\left (\protect \pazocal {N}(\mathinner {|{\alpha /\protect \sqrt {n_1
  \times n_2}}\rangle }\protect \!\mathinner {\langle {\alpha /\protect \sqrt {n_1 \times n_2}}|})\right )^{\boxplus n_1}\right )^{\boxplus n_2} \\ &= \protect \pazocal {N}(\mathinner {|{\alpha /\protect \sqrt {n_1 \times n_2}}\rangle }\protect \!\mathinner {\langle {\alpha /\protect \sqrt {n_1 \times n_2}}|})^{\boxplus (n_1 \times n_2)}. \end {aligned}\end {equation}Following the fact that the span of coherent states are dense in the Banach space of trace class operators, we can conclude that $\protect \pazocal {N}^{\boxplus (n_1 \times n_2)}= \left (\protect \pazocal {N}^{\boxplus n_1}\right )^{\boxplus n_2}$.

\bibitem{winter2017energy}
Andreas Winter.
\newblock Energy-constrained diamond norm with applications to the uniform continuity of continuous variable channel capacities, 2017.
\newblock URL: \url{https://arxiv.org/abs/1712.10267}, \href {https://arxiv.org/abs/1712.10267} {\path{arXiv:1712.10267}}.

\bibitem{lami2023exact}
Ludovico Lami and Mark~M Wilde.
\newblock Exact solution for the quantum and private capacities of bosonic dephasing channels.
\newblock {\em Nature Photonics}, 17(6):525--530, 2023.
\newblock \href {https://doi.org/10.1038/s41566-023-01190-4} {\path{doi:10.1038/s41566-023-01190-4}}.

\bibitem{Serafini}
Alessio Serafini.
\newblock {\em Quantum continuous variables: a primer of theoretical methods}.
\newblock CRC press, 2017.
\newblock \href {https://doi.org/10.1201/9781315118727} {\path{doi:10.1201/9781315118727}}.

\bibitem{Note2}
Let $\protect \mathrm {HS}$ denote the Hilbert space of trace-class operators with finite Hilbert-Schmidt norm, i.e., those $T$ satisfying $\protect \Tr \left [ T^\dagger T\right ] < +\infty $. Finite rank operators belong to $\protect \mathrm {HS}$. Thus, it suffices to show that the span of coherent states is dense in $\protect \mathrm {HS}$. Let $V$ be the linear span of coherent states. If $V$ were not dense in $\protect \mathrm {HS}$, then there would exist a nonzero operator $T \in \protect \mathrm {HS}$ orthogonal to $V$, meaning $\mathinner {\langle {\alpha }|} T \mathinner {|{\alpha }\rangle } = 0$ for all $\alpha \in \protect \mathds {C}$. This implies that the Husimi function of $T$ vanishes identically, which forces $T$ to be the zero operator~\cite {Serafini}.

\end{thebibliography}

\appendix
\section{Proof of Theorem~\ref{Thm:qCLTChannel}}
In this section, we present the proof of our main result, the qCLT for bosonic quantum channels stated in Theorem~\ref{Thm:qCLTChannel}. Our goal is to show that, for every trace-class operator $T$, the sequence $\NN^{\boxplus n}(T)$ converges to $\NN_G(T)$ as $n$ goes to infinity. It is well known that the span of coherent states $\ketbra{\alpha}$ are dense in the Banach space of trace-class operators $\T(\HH)$~\footnote{Let $\mathrm{HS}$ denote the Hilbert space of trace-class operators with finite Hilbert-Schmidt norm, i.e., those $T$ satisfying $\Tr \left[ T^\dagger T\right] < +\infty$. Finite rank operators belong to $\mathrm{HS}$. Thus, it suffices to show that the span of coherent states is dense in $\mathrm{HS}$. Let $V$ be the linear span of coherent states. If $V$ were not dense in $\mathrm{HS}$, then there would exist a nonzero operator $T \in \mathrm{HS}$ orthogonal to $V$, meaning $\bra{\alpha} T \ket{\alpha} = 0$ for all $\alpha \in \C$. This implies that the Husimi function of $T$ vanishes identically, which forces $T$ to be the zero operator~\cite{Serafini}.}. Moreover, since both $\NN^{\boxplus n}$ and $\NN_G$ are completely-positive trace-preserving maps, we have 
\bb\label{eq:ContractiveTraceNorm}
	\| \NN^{\boxplus n}(T) \|_1 \leq \| T \|_1, \quad \| \NN_G(T) \|_1 \leq \| T \|_1,
\ee
where $\| T \|_1 \coloneqq \Tr \left[ \sqrt{T^\dagger T} \right]$ is the trace-norm. The above equation asserts that $\NN^{\boxplus n}$ and $\NN_G$ have uniformly bounded operator norm. Consequently, to prove the theorem it is sufficient to show that, for every $\alpha \in \C$, the sequence $\NN^{\boxplus n}(\ketbra{\alpha})$ converges to $\NN_G(\ketbra{\alpha})$ in trace norm as $n$ tends to infinity.

To do so, we begin with the so-called \emph{Strong-Weak Operator Topology} (SWOT) Lemma (see, e.g.,~\cite[Lemma 4]{lami2018all}), which allows us to reduce convergence of the sequence $\NN^{\boxplus n}(\ketbra{\alpha})$ to $\NN_G(\ketbra{\alpha})$ in trace-norm to the pointwise convergence of their characteristic functions. Thus, by the SWOT Lemma, it is sufficient to show that $$\lim_{n \to \infty}\chi_{\NN^{\boxplus n}(\ketbra{\alpha})}(z) = \chi_{\NN_G(\ketbra{\alpha})}(z),$$for every $z \in \C$.

To show this, we first use the action of the $n$-beam splitter unitary on displacement operators given in~\eqref{eq:def-U-n}, from which we obtain
\bb
	U_n \left(\ket{\alpha_1} \otimes \cdots \otimes \ket{\alpha_n} \right) = \left(\ket{\beta_1} \otimes \cdots \otimes \ket{\beta_n} \right),
\ee
where $(\beta_1, \ldots, \beta_n)^\top \coloneqq V_n (\alpha_1, \ldots, \alpha_n)^\top$. As a result, we can express the action of $\NN^{\boxplus n}$ on coherent states as
\bb\label{eq:SymmetricConvolutionCoherent}
&\NN^{\boxplus n} (\ketbra{\alpha})\\
&= \Tr_{\neg 1} \! \left[U_n\, \NN^{\otimes n} \left(U_n^\dag\, \left(\ketbra{\alpha}\otimes \ketbra{0}^{\otimes (n-1)}\right) \, U_n \right) U_n^\dag \right] \\
&= \Tr_{\neg 1} \! \left[U_n\, \NN^{\otimes n} \left(\Ketbra{\alpha/\sqrt{n}}^{\otimes n} \right) U_n^\dag \right] \\
&= \NN \left(\Ketbra{\alpha/\sqrt{n}} \right)^{\boxplus n}\, ,
\ee
for every $\alpha \in \C$. Following the above equation, and using once again the action of the $n$-beam splitter unitary on displacement operators, we obtain the following expression at the level of characteristic functions:
\bb\label{eq:CharConvCoherent}
	\chi_{\NN \left(\Ketbra{\alpha/\sqrt{n}} \right)^{\boxplus n}}\left(z\right) = \chi_{\NN \left(\Ketbra{\alpha/\sqrt{n}} \right)}\left( \frac{z}{\sqrt n}\right)^n.
\ee
The above equation suggests that, as $n$ tends to infinity, we must understand the behavior of the channel $\NN$ acting on coherent states $\ket{\alpha}$ with relatively small $|\alpha|$, as well as the behavior of its characteristic function near the origin.

For coherent states $\ket{\alpha}$ with small $|\alpha|$, we can take advantage of their Fock-basis expansion $\ket{\alpha} = e^{-|\alpha|^2/2} \sum_{n \in \Z^{\geq 0}} \frac{\alpha^n}{\sqrt{n!}}\, \ket{n},$ which yields the following expression for small~$|\alpha|$:
\bb\label{eq:approximationCoherent}
&\ketbra{\alpha} = \left(1- |\alpha|^2\right)\ketbra{0} + \alpha \ketbraa{1}{0} + \alpha^* \ketbraa{0}{1} \\
&+ \frac{\alpha^2}{\sqrt2} \ketbraa{2}{0} + |\alpha|^2 \ketbra{1} + \frac{{\alpha^*}^2}{\sqrt2} \ketbraa{0}{2} + o\big(|\alpha|^2\big)\,.
\ee
Here the error term $o\big(|\alpha|^2\big)$ is understood with respect to the trace distance.

Moreover, for the behavior of the characteristic function near the origin, we can relate that of a trace-class operator $T$ to its moments via a Taylor expansion. In particular, we take advantage of the following useful relations
\bb
\partial_{z} \chi_{T}(z) = \chi_{\frac 12(\ac^{\dagger}T +T\ac^\dagger)}(z), \quad \partial_{\bar z} \chi_{T}(z) = -\chi_{\frac 12(\ac T +T\ac)}(z).
\ee

To proceed, note that by the assumptions of Theorem~\ref{Thm:qCLTChannel}, the state $\NN(\ketbra{0})$ is centered and has a finite second-order moment. Consequently, its characteristic function is twice differentiable at the origin, and we may write the following Taylor expansion in a neighborhood of the origin:
\bb\label{eq:TaylorVacuum}
	&\chi_{\NN(\ketbra{0})}(z) \\
	&= 1 + \frac12 \Big( z^2 \Tr\left[\NN(\ketbra{0}) {\ac^\dagger}^2\right] + {\bar z}^2 \Tr\left[\NN(\ketbra{0}) \ac^2 \right]\\
	&\qquad - |z|^2 \Tr \left[  \NN(\ketbra{0}) \left( 2 \ac^\dagger \ac + 1\right)\right]  \Big) + o\big( |z|^2 \big).
\ee
Also, again by assumption, we know that $\NN(\ketbraa{0}{1})$ and $\NN({\ketbraa{1}{0}})$ are trace-less operators with finite first-order moments. Thus, in a neighborhood of the origin, we also have the expansions 
\bb\label{eq:TaylorOne}
&\chi_{\NN(\ket{0}\!\bra{1})}(z)\\
& \quad = z \Tr \left[ \NN(\ket{0}\!\bra{1}) \ac^\dagger \right] - \bar z \Tr \left[ \NN(\ket{0}\!\bra{1}) \ac \right] + o(|z|)\, , \\
&\chi_{\NN(\ket{1}\!\bra{0})}(z) \\
&\quad =  z \Tr \left[ \NN(\ket{1}\!\bra{0}) \ac^\dagger \right] - \bar z \Tr \left[ \NN(\ket{1}\!\bra{0}) \ac \right] + o(|z|)\, .
\ee

Starting from the approximation of coherent states given in~\eqref{eq:approximationCoherent}, we can write
\begin{widetext}
\bb\label{eq:Characteristic_Convolution_Coherent}
&\chi_{\NN\left(\Ketbra{\alpha/\sqrt{n}}\right)}\big(z/\sqrt{n}\big) \\
&\quad = \left(1- \frac{|\alpha|^2}{n} \right) \chi_{\NN(\ket{0}\!\bra{0})}\big(z/\sqrt{n}\big) + \frac{\alpha}{\sqrt{n}}\,\chi_{\NN( \ket{1}\!\bra{0})}\big(z/\sqrt{n}\big) \\
&\qquad + \frac{\alpha^*}{\sqrt{n}}\, \chi_{\NN(\ket{0}\!\bra{1})}\big(z/\sqrt{n}\big) + \frac{1}{\sqrt2\, n}\, \chi_{\NN\left( \alpha^2 \ket{2}\!\bra{0} + \sqrt2 |\alpha|^2 \ket{1}\!\bra{1} + {\alpha^*}^2 \ket{0}\!\bra{2}\right)}\big(z/\sqrt{n}\big) + o\big(1/n\big) \\
&\quad = 1 + \frac{1}{2n} \left( z^2 \Tr\left[\NN(\ketbra{0}) {\ac^\dagger}^2\right] + {\bar z}^2 \Tr\left[\NN(\ketbra{0}) \ac^2 \right] - |z|^2 \Tr \left[  \NN(\ketbra{0}) \left( 2 \ac^\dagger \ac + 1\right)\right]  \right)  \\
&\qquad -\frac{|\alpha|^2}{n} + \frac{\alpha}{n} \left( z \Tr \left[ \NN(\ket{0}\!\bra{1}) \ac^\dagger \right] - \bar z \Tr \left[ \NN(\ket{0}\!\bra{1}) \ac \right] \right) + \frac{\alpha^*}{n} \left(z \Tr \left[ \NN(\ket{1}\!\bra{0}) \ac^\dagger \right] - \bar z \Tr \left[ \NN(\ket{1}\!\bra{0}) \ac \right] \right) + \frac{|\alpha|^2}{n} + o\big(1/n\big) \\
&\quad = 1 + \frac{1}{2n} \left( z^2 \Tr\left[\NN(\ketbra{0}) {\ac^\dagger}^2\right] + {\bar z}^2 \Tr\left[\NN(\ketbra{0}) \ac^2 \right] - |z|^2 \Tr \left[  \NN(\ketbra{0}) \left( 2 \ac^\dagger \ac + 1\right)\right]  \right)  \\
&\qquad + \frac{1}{n} \left( \alpha z \Tr \left[ \NN(\ket{0}\!\bra{1}) \ac^\dagger \right] - \alpha \bar z \Tr \left[ \NN(\ket{0}\!\bra{1}) \ac \right] + \alpha^* z \Tr \left[ \NN(\ket{1}\!\bra{0}) \ac^\dagger \right] - \alpha^* \bar z \Tr \left[ \NN(\ket{1}\!\bra{0}) \ac \right] \right) + o\big(1/n\big) \, ,
\ee
\end{widetext}
The second line follows from~\eqref{eq:TaylorVacuum} and~\eqref{eq:TaylorOne}, together with the continuity of the characteristic function for trace-class operators, namely that $\chi_T(z) = \Tr \left[ T\right] + o(1)$ in a neighborhood of the origin, to handle the characteristic function of $\NN\left( \alpha^2 \ket{2}\!\bra{0} + \sqrt2 |\alpha|^2 \ket{1}\!\bra{1} + {\alpha^*}^2 \ket{0}\!\bra{2}\right)$. Now, using the expansions derived in~\eqref{eq:Characteristic_Convolution_Coherent} and the relation~\eqref{eq:CharConvCoherent}, we obtain
\bb\label{eq:GaussificationCorherent}
&\lim_{n\to\infty} \chi_{\NN^{\boxplus n} (\ket{\alpha}\!\bra{\alpha})}(z)  =\\
&\exp\!\left[ \frac{\left( z^2 G -|z|^2 H + {\bar z}^2 G^* \right)}{2} + z (\alpha t^* + \alpha^* s^*) - \bar z (\alpha s + \alpha^* t) \right],
\ee
where here
\begin{align}
t \coloneqq \Tr \left[ \NN(\ket{1}\!\bra{0}) \ac \right]&, \quad s \coloneqq \Tr \left[ \NN(\ket{0}\!\bra{1}) \ac \right],\\
G \coloneqq \Tr\left[\NN(\ketbra{0}) {\ac^\dagger}^2\right]&, \quad H  \coloneqq \Tr \left[  \NN(\ketbra{0}) \left( 2 \ac^\dagger \ac + 1\right)\right].
\end{align}
For the final step, note that $\ketbra{\alpha}$ is a Gaussian quantum state with $2\times2$ identity covariance matrix and first-order moment vector $\sqrt 2 \left( \Re \alpha, \Im \alpha\right)^\top$. Using the action of the Gaussian channel $\NN_G$, with parameters given in~\eqref{eq:XqCLT}, on first- and second-order moments as described in~\eqref{eq:GauusianChannel_Covariance}, together with the characteristic function of Gaussian states in~\eqref{eq:CharGaussian}, we can simply verify that $$\lim_{n\to\infty} \chi_{\NN^{\boxplus n} (\ket{\alpha}\!\bra{\alpha})}(z)  =  \chi_{\NN_G (\ket{\alpha}\!\bra{\alpha})}(z)$$ holds for every $z \in \C$.

\end{document}